\documentclass[runningheads]{llncs}
\usepackage{graphicx}
\usepackage{url}            
\usepackage{booktabs}       
\usepackage{amsfonts}       
\usepackage{nicefrac}       
\usepackage{microtype}      
\usepackage{tikz}
\usepackage{tabularx}
\usepackage{hhline}
\usepackage{adjustbox}
\usepackage{multirow}
\usepackage{mathtools}
\usepackage{bbm}
\usepackage{caption}
\usepackage{cellspace}
\usepackage{xcolor}
\usepackage{hyperref}
\usepackage{marvosym}
\DeclarePairedDelimiterX\set[1]\{\}{\nonscript\,#1\nonscript\,}

\usetikzlibrary{shapes.geometric, arrows}
\tikzstyle{startstop} = [rectangle, rounded corners, minimum width=1.5cm, minimum height=0.8cm,text centered, draw=black, fill=red!30]
\tikzstyle{io} = [trapezium, trapezium left angle=70, trapezium right angle=110, minimum width=1.5cm, minimum height=0.8cm, text centered, draw=black, fill=blue!30]
\tikzstyle{process} = [rectangle, minimum width=1.5cm, minimum height=0.8cm, text centered, draw=black, fill=orange!30]
\tikzstyle{decision} = [diamond, minimum width=1.5cm, minimum height=0.8cm, text centered, draw=black, fill=green!30]
\tikzstyle{block} = [rectangle, rounded corners, minimum width=1.5cm, minimum height=0.8cm,text centered, draw=black, fill=orange!30]
\tikzstyle{network} = [rectangle, rounded corners, minimum width=1.5cm, minimum height=0.8cm,text centered, draw=black, fill=green!30]
\tikzstyle{bim} = [rectangle, rounded corners, minimum width=0.1cm, minimum height=0.1cm, text centered, draw=blue, dash pattern=on 4pt off 4pt]
\tikzstyle{arrow} = [thick,->,>=stealth]

\DeclareMathOperator*{\argmax}{arg\,max}
\DeclareMathOperator*{\argmin}{arg\,min}

\newcommand{\dtoprule}{\specialrule{1pt}{0pt}{1.5pt}%
            \specialrule{0.3pt}{0pt}{\belowrulesep}%
            }
\newcommand{\dbottomrule}{\specialrule{0.3pt}{0pt}{1.5pt}%
            \specialrule{1pt}{0pt}{\belowrulesep}%
            }
\newcommand{\dmidrule}{\specialrule{0.3pt}{0pt}{1.5pt}%
            \specialrule{0.3pt}{0pt}{\belowrulesep}%
            }

\newcommand{\ddtoprule}{\specialrule{1pt}{0pt}{1.5pt}%
            \specialrule{0.3pt}{0pt}{0pt}%
            }

\newcommand{\ddmidrule}{\specialrule{0.3pt}{0pt}{1.5pt}%
            \specialrule{0.3pt}{0pt}{0pt}%
            }

\usepackage{makecell}

\definecolor{lime}{HTML}{A6CE39}
\DeclareRobustCommand{\orcidicon}{%
	\begin{tikzpicture}
	\draw[lime, fill=lime] (0,0) 
	circle [radius=0.16] 
	node[white] {{\fontfamily{qag}\selectfont \tiny ID}};
	\draw[white, fill=white] (-0.0625,0.095) 
	circle [radius=0.007];
	\end{tikzpicture}
	\hspace{-2mm}
}

\foreach \x in {A, ..., Z}{%
	\expandafter\xdef\csname orcid\x\endcsname{\noexpand\href{https://orcid.org/\csname orcidauthor\x\endcsname}{\noexpand\orcidicon}}
}

\newcommand\hlb[1]{\textcolor{black!80!black}{#1}}

\usepackage{float}
\floatstyle{plaintop}
\restylefloat{table}

\begin{document}
\title{Learning correspondences of cardiac motion from images using biomechanics-informed modeling}
\titlerunning{Learning spatial-temporal correspondences of cardiac motion}
%
\author{Xiaoran Zhang\inst{1}\textsuperscript{(\Letter)}\orcidA{} \and Chenyu You\inst{2} \and Shawn Ahn\inst{1}\orcidB{} \and Juntang Zhuang\inst{1} \and Lawrence Staib\inst{1,2,3}\orcidC{} \and James Duncan\inst{1,2,3}\orcidD{}}
\authorrunning{Zhang et al.}
%
\institute{Biomedical Engineering, Yale University, New Haven CT, USA \\\email{xiaoran.zhang@yale.edu} \and Electrical Engineering, Yale University, New Haven CT, USA \and Radiology \& Biomedical Engineering, Yale School of Medicine, New Haven CT, USA}
%
\maketitle              
\begin{abstract}
Learning spatial-temporal correspondences in cardiac motion from images is important for understanding the underlying dynamics of cardiac anatomical structures. Many methods explicitly impose smoothness constraints such as the $\mathcal{L}_2$ norm on the displacement vector field (DVF), while usually ignoring biomechanical feasibility in the transformation. Other geometric constraints either regularize specific regions of interest such as imposing incompressibility on the myocardium or introduce additional steps such as training a separate network-based regularizer on physically simulated datasets. In this work, we propose an explicit biomechanics-informed prior as regularization on the predicted DVF in modeling a more generic biomechanically plausible transformation within all cardiac structures without introducing additional training complexity. We validate our methods on two publicly available datasets in the context of 2D 
MRI data and perform extensive experiments to illustrate the effectiveness and robustness of our proposed methods compared to other competing regularization schemes. Our proposed methods better preserve biomechanical properties by visual assessment and show advantages in segmentation performance using quantitative evaluation metrics. The code is publicly available at \url{https://github.com/Voldemort108X/bioinformed_reg}. 
\end{abstract}

\section{Introduction}
A displacement vector field (DVF) estimated by medical image registration is crucial to infer the underlying spatial-temporal characteristics of anatomical structures. The DVF is especially important for cardiac motion, whose functional assessment is often analyzed using cine MRI or echocardiography \cite{zhang2021fully,ahn2021multi,zhang2018fully,zhang2020fully}. 
Specifically, accurate prediction of the DVF across cardiac sequences enables regional myocardial strain estimation, which contributes to the localization of myocardial infarction \cite{lu2021learning}.  

A wide range of deformable image registration methods have been studied for medical images to obtain reliable DVF measurement. Regularization is often imposed to ensure smoothness along with a dissimilarity measure. Such methods include B-splines \cite{rueckert1999nonrigid}, which use the bending energy of a thin-plate of metal, and optical flow \cite{perez2013tv}, which uses total variation. Diffeomorphism is also a commonly used regularization assumption, which parameterizes the DVF as a set of velocity fields to ensure invertibility \cite{ashburner2007fast}. For cardiac motion estimation, incompressibility of the myocardium is also used to incorporate physiology-inspired prior into the registration process by enforcing, for example, a divergence-free constraint \hlb{\cite{gao2021optimization}}.

With the advent of deep learning, convolutional neural networks have been widely applied to unsupervised learning of deformable registration and segmentation \cite{you2021momentum,you2022class,zhang2021automatic,you2020unsupervised,li2022myops,you2022simcvd,you2022bootstrapping,you2022incremental}. Diffeomorphic constraints \cite{dalca2018unsupervised} and $\mathcal{L}_2$ norms on the displacement gradient \cite{balakrishnan2019voxelmorph} are usually used for regularization. However, biomechanical plausibility is usually not explicitly considered and thus the estimated displacement field might result in unrealistic motion. To encourage physically plausible transformation, pre-trained variational auto-encoders (VAEs) can be used as regularizers in the registration pipeline to enforce implicit constraints on the DVF \cite{qin2020biomechanics,sang2020enhanced,sang20214d}. However, such VAE regularizers are often trained separately on manually generated datasets which require physical simulations and thus can be time-consuming and complicated. Due to the intra- and inter-variability in clinical datasets, such network-based regularizers might also be biased when evaluating new datasets. In addition, explicit biomechanical constraints such as incompressibility are often considered 
only for specific regions of interest, such as the myocardium \cite{qin2020biomechanics,ahn2020unsupervised} for cardiac applications, while the preservation of geometric properties in other structures such as the right ventricle (RV) is not guaranteed.

In this work, we propose a biomechanics-informed regularization explicitly as a prior for DVF estimation to better preserve geometric properties \hlb{without requiring myocardium segmentation}. Using a separate network as a regularizer might introduce additional steps and thus increase complexity and computation. By contrast, our proposed methods show advantages in generalization when applying to new datasets. The consistency across different cardiac structures is also improved in our proposed methods due to the more generic assumption. We validate our methods on two public datasets and conduct extensive experiments to show its effectiveness and robustness over other competing regularization techniques.

\section{Method}

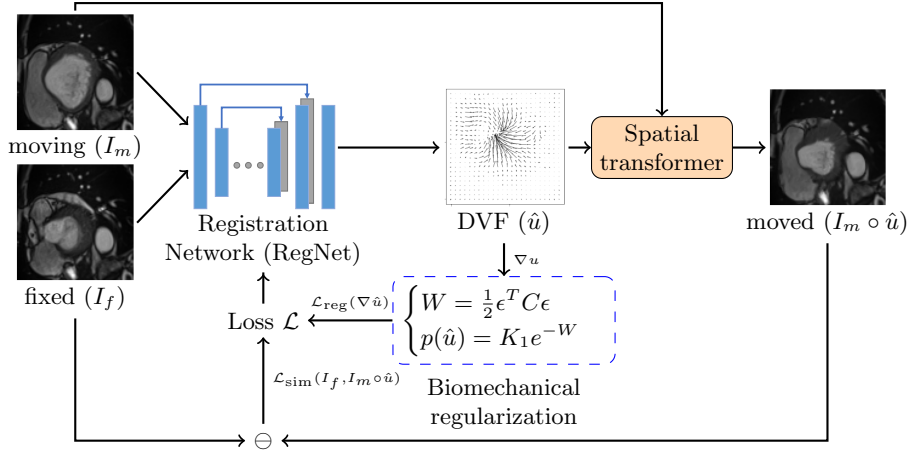
\begin{figure}[htbp]
        \centering
        \begin{tikzpicture}[align=center]
            \node[inner sep=0pt] (mov) at (0,-1) {\includegraphics[scale=0.2]{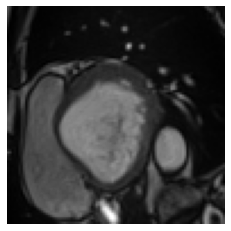}};
            \node (movtext) [below of=mov] {\hlb{moving ($I_m$)}};
            \node[inner sep=0pt] (fix) at (0,-3) {\includegraphics[scale=0.2]{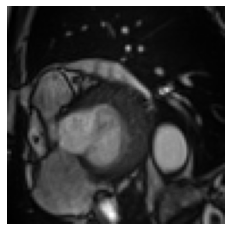}};
            \node (fixtext) [below of=fix] {\hlb{fixed ($I_f$)}};
            \node[inner sep=0pt] (nn) at (2.5,-2) {\includegraphics[scale=0.3]{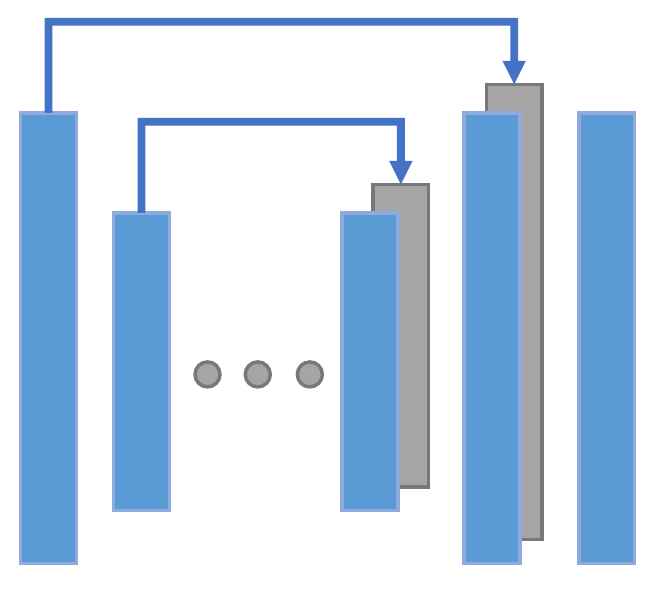}} ;
            \node (nntext) [below of=nn, yshift=-0.23cm] {Registration \\Network (RegNet)};
            \node (loss) [below of=nn, yshift=-1.3cm] {Loss $\mathcal{L}$};
            \node (minus) [below of=loss, yshift=-0.6cm] {$\ominus$};
            \node[inner sep=0pt] (dvf) at (5.7,-2) {\includegraphics[scale=0.2]{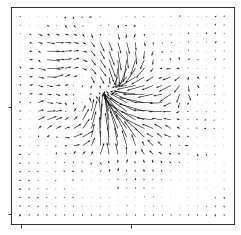}};
            \node (stn) [block, right of=dvf, xshift=1.1cm] {Spatial \\ transformer};
            \node (bim) [bim, below of=dvf, yshift=-1.3cm] {$\begin{cases}
            \hlb{W = \frac{1}{2}\epsilon^T C\epsilon} \\
            p(\hat{u}) = K_1e^{-W}
            \end{cases}$};
            \node[inner sep=0pt] (bimtext) [below of=bim, yshift=-0.1cm] {Biomechanical \\ regularization};
            \node (dvftext) [below of=dvf] {\hlb{DVF ($\hat{u}$)}};
            \node[inner sep=0pt] (moved) at (10,-2) {\includegraphics[scale=0.2]{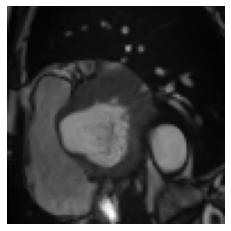}};
            \node (movedtext) [below of=moved] {\hlb{moved ($I_m\circ \hat{u}$)}};
            
            \draw[->,thick] (minus.north) --node[anchor=west]{$\scriptscriptstyle\mathcal{L}_{\text{sim}}(I_f, I_m\circ \hat{u})$} (loss.south);
            \draw[->,thick] (loss.north) |- (nntext.south);
            \draw[->,thick] (fixtext.south) |- (minus.west);
            \draw[->,thick] (mov.east) -- (1.5, -1.7);
            \draw[->,thick] (fix.east) -- (1.5, -2.3);
            \draw[->,thick] (nn.east) -- (dvf.west);
            \draw[->,thick] (dvf.east) -- (stn.west);
            \draw[->,thick] (stn.east) -- (moved.west);
            \draw[->,thick] (mov.north) -- ++(0,0.1) -| (stn.north);
            \draw[->,thick] (movedtext.south) |- (minus.east);
            \draw[->,thick] (bim.west) --node[anchor=south]{$\scriptscriptstyle \mathcal{L}_{\text{reg}}(\nabla \hat{u})$} (loss.east);
            \draw[->,thick] (dvftext.south) --node[anchor=west]{$\scriptscriptstyle\nabla u$} (bim.north);
        \end{tikzpicture}
        \caption{Illustration of the proposed framework using biomechanics-informed modeling (BIM) for cardiac image registration. 
        }
        \label{fig:framework}
    \end{figure}

The proposed framework is illustrated in Fig.~\ref{fig:framework}. We utilize an explicit biomechanical regularization on the estimated displacement vector field (DVF) from the registration network (RegNet), which takes pairs of moving and fixed images as inputs $(I_m, I_f)\in \mathbbm{R}^{2\times C\times H\times W}$ and outputs the estimated DVF, $\hat{u}\in \mathbbm{R}^{2\times H\times W}$. The following sections describe a detailed biomechanics-informed modeling (BIM) process (Sec. 2.1) and the derivation of the proposed loss function (Sec. 2.2).

\subsection{Biomechanics-informed modeling}
From the classical definition of the strain tensor $\epsilon\in\mathbbm{R}^{3\times H\times W}$ (defined in Eq.~\ref{eq:sigma}) 
in the infinitesimal linear elasticity model \cite{spencer2004continuum} for 2D displacement $\hat{u}=[\hat{u}_1, \hat{u}_2]^T$ 
\hlb{\begin{equation}
    \epsilon = \left[\frac{\partial \hat{u}_1}{\partial x_1},  \frac{\partial \hat{u}_2}{\partial x_2}, \frac{1}{2}\left(\frac{\partial \hat{u}_1}{\partial x_2} + \frac{\partial \hat{u}_2}{\partial x_1}\right) \right]^T, \label{eq:sigma}
\end{equation}}
we define the \hlb{linear isotropic elastic strain energy density function for each pixel} as follows:
\hlb{\begin{equation}
    W_{i,j} = \frac{1}{2}\epsilon_{i,j}^TC\epsilon_{i,j},
\end{equation}}
where $C$ is the stiffness matrix describing material properties of the deforming body
\hlb{\begin{equation}
    C^{-1} = \begin{bmatrix}
        \frac{1}{E_p} & \frac{-\nu_p}{E_p} & 0 \\
        \frac{-\nu_p}{E_p} & \frac{1}{E_p} & 0 \\
        0 & 0 & \frac{2(1+\nu_p)}{E_p}
    \end{bmatrix}.
\end{equation}}
\hlb{$E_p$ is defined as stiffness and $\nu_p$ as Poisson ratio \cite{papademetris2002estimation}.} In this paper, we treat $E_p$ as constant and $\nu_p$ as a hyperparameter. The prior probability density function (pdf) of the DVF can be written in Gibb's form:
\begin{equation}
    p(\hat{u}) = k_1e^{-W} \label{eq:prior}.
\end{equation}
The optimal DVF $\hat{u}^*$ can be obtained through maximum a posteriori (MAP) optimization:
\begin{align}
    \hat{u}^* &= \underset{\hat{u}}{\argmax}\; \left\{ p(\hat{u}|u) = \frac{p(u|\hat{u})p(\hat{u})}{p(u)} \right\} 
        = \underset{\hat{u}}{\argmin}\; \left\{ -\log p(u|\hat{u}) -\log p(\hat{u}) \right\}.
\end{align}

\subsection{Loss function}
We assume the noise between the ground truth measurement $u$ and the DVF estimate $\hat{u}$ is normally distributed $\mathcal{N}(0, \sigma^2)$ 
\begin{equation}
    p(u|\hat{u}) = \frac{1}{\sqrt{2\pi}\sigma}e^{-\frac{||u-\hat{u}||_2^2}{2\sigma^2}}.
\end{equation}
As the ground truth measurement $u$ is usually hard to acquire, we utilize the image dissimilarity between the transformed image $I_m\circ \hat{u}$ and the reference image $I_f$ to evaluate the difference of $\hat{u}$ with the ground truth motion field. From the prior pdf of DVF in Eq.~\ref{eq:prior}, our proposed loss can be written as:
\hlb{\begin{align}
    \mathcal{L}(I_m, I_f; \hat{u}) = \underbrace{\frac{1}{N}\sum_{i=1}^N ||I_f^i-I_m^i\circ \hat{u}^i||_2^2}_{\mathcal{L}_{sim}} + \lambda \underbrace{\frac{1}{N}\sum_{i=1}^N||\epsilon^T C\epsilon||_2}_{\mathcal{L}_{reg}},
\end{align}}
where $N$ is the number of samples. \hlb{We apply $\mathcal{L}_2$ norm for the image after computing the strain energy density with scalar-valued entry for each pixel to penalize unrealistic motion.} When segmentation masks are available for certain cardiac structures for an input image pair $(I_m, I_f)$, we can further improve the approximated difference of $\hat{u}$ with ground truth previously computed using image dissimilarity by adding an auxiliary segmentation loss term
\begin{align}
    \mathcal{L}(I_m, I_f, s_m, s_f; \hat{u}) = \mathcal{L}(I_m, I_f; \hat{u}) + \gamma \underbrace{\frac{1}{N}\sum_{i=1}^{N}\sum_{j=1}^K\left(1-\text{Dice}\left(s_m^{ij}\circ \hat{u}^i, s_f^{ij} \right)\right)}_{\mathcal{L}_{seg}},
\end{align}
where $(s_m^{ij}, s_f^{ij})$ denotes the segmentation mask of cardiac structure $j$ for image pair $(I_m^i, I_f^i)$ with sample index $i$ and $j=1,2,...,K$.

\section{Experiments}
\subsection{Dataset details}
We validate the effectiveness of our proposed methods on the publically-available ACDC 2017 dataset \cite{bernard2018deep}. 
It contains 100 patients including the following five categories with even split: 1) healthy, 2) previous myocardial infarction, 3) dilated cardiomyopathy, 4) hypertrophic cardiomyopathy, and 5) abnormal right ventricle (RV). Segmentation labels of left ventricle (LV), RV, epicardium (Epi) boundaries for end-diastolic (ED) and end-systolic (ES) frames are given in the dataset. [For both datasets, indicate that they are cine MR and the mm dimensions] For this data, we use 60 patients for training, 20 for validation, and 20 for testing, after a random shuffle. 

We also validate our method on another publicly available LV quantification dataset from 2019 
\cite{xue2018full}. It contains 56 subjects and includes 20 frames of the entire cardiac cycle. Segmentation labels of the endocardium (Endo) and epicardium boundaries are given for each frame in the cycle. In this paper, 34 patients are used for training, 11 for validation, and 11 for testing, after a random shuffle.

\subsection{Experimental setup}
\hlb{We set the ED frame as the moving image and the ES frame as the fixed image in the ACDC and LV quantification datasets in order to estimate cardiac contraction}. Each slice in both datasets is cropped to $96\times 96$ with respect to the myocardial centroid after min-max normalization. We compare our proposed framework with several state-of-the-art registration methods including 1) B-splines \cite{rueckert1999nonrigid} implemented using SimpleElastix, 2) optical flow \cite{perez2013tv} implemented using scikit-image registration package, 3) a learning-based registration framework using VAE regularization (RegNet+VAE) \cite{qin2020biomechanics}, and 4) a learning-based registration framework using $\mathcal{L}_2$ regularization (RegNet+$\mathcal{L}_2$) \cite{balakrishnan2019voxelmorph}. The learning-based registration algorithms are implemented using Pytorch with Adam optimizer on 100 epochs. The learning-based models run 18-30 minutes on a single NVIDIA GTX 2080Ti GPU with 12 GB memory for training and validation. The loss curves in training and validation are shown in the supplementary material (Fig.~\ref{fig:loss_curve}). We choose $\nu_p=0.4$ (Poisson ratio), $\lambda=0.05$ (weight for BIM loss $\mathcal{L}_{\text{reg}}$), and $\gamma=0.01$ (weight for auxiliary segmentation loss $\mathcal{L}_{\text{seg}}$) in our proposed methods after optimizing hyperparameters. \hlb{The hyperparameters of the competitor models are optimized as well.}

\subsection{Evaluation metrics}
To evaluate our proposed method, we compute the Dice coefficient (DC), Jaccard index (JD), Hausdorff distance (HD), and average symmetric surface distance (ASD) to evaluate segmentation conformance. We also compute the Jacobian determinant to evaluate the quality of the generated DVF.

\section{Results}
\subsection{Visual and quantitative assessments}
\begin{figure}
    \centering
    \includegraphics[scale=0.175]{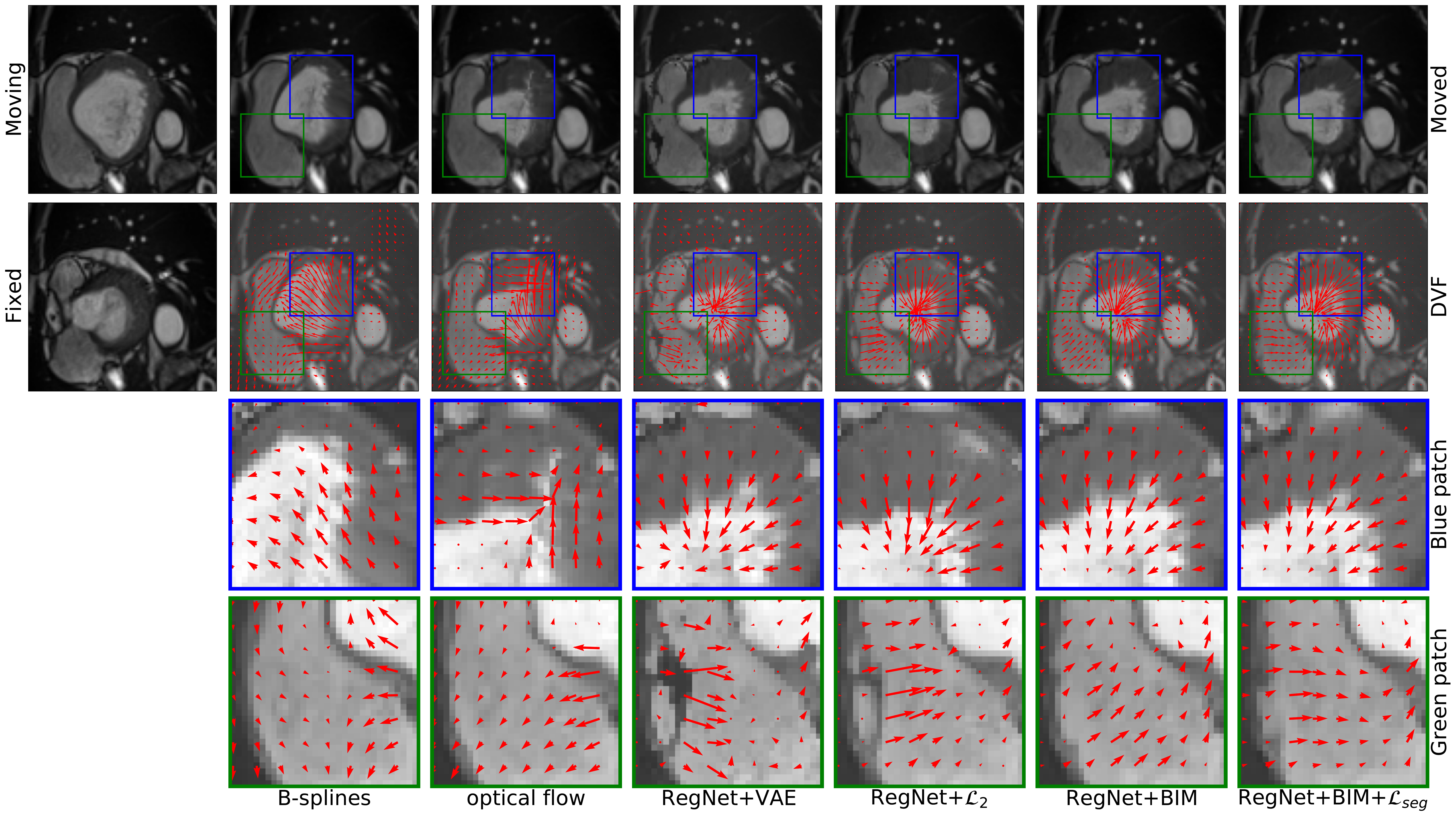}
    \caption{Visual assessment of registration performance 
    on the ACDC 2017 dataset \cite{bernard2018deep}. 
    }
    \label{fig:ACDC_visual}
\end{figure}

Visual assessment of our proposed methods compared to other approaches on the ACDC dataset is shown in Fig.~\ref{fig:ACDC_visual}. From the figure, we can see that B-splines \cite{rueckert1999nonrigid} and optical flow \cite{perez2013tv} generate artifacts in the myocardium, especially in the highlighted blue patch shown in the third row. The RegNet+VAE \cite{qin2020biomechanics} creates unrealistic motion in the right ventricle region, especially in the highlighted green patch shown in the last row, which might be due to penalization of the VAE regularizer mainly focusing on the myocardium only. Our proposed methods, in the last two columns, show improved motion compared to the RegNet+$\mathcal{L}_2$ with smoother and more biomechanically plausible motion that preserves the geometric features of cardiac structures. The visual results of our proposed methods compared with others on the LV quantification dataset is included in the supplementary material (Fig.~\ref{fig:LVQuant_visual}).

\begin{table}[htbp]
    \centering
    \begin{adjustbox}{scale=0.9}
    \caption{Quantitative assessment of registration performance 
    on the ACDC dataset \cite{bernard2018deep}.}
    \begin{tabular}{llccccc}
    \dtoprule
                     &   Method  & Dice[\%]$\uparrow$    & Jaccard[\%]$\uparrow$    & \hlb{HD[mm]}$\downarrow$      & \hlb{ASD[mm]}$\downarrow$\\\midrule
    \multirow{7}{*}{LV} & affine only & 72.01 & 57.24 & \hlb{11.77} & \hlb{2.13}\\
                             & B-splines \cite{rueckert1999nonrigid} & 89.96 & 82.62 & \hlb{5.37} & \hlb{0.48} \\
                             & optical flow \cite{perez2013tv} & 87.07 & 77.87 & \hlb{7.65} & \hlb{0.74}\\ 
                             & RegNet+VAE \cite{qin2020biomechanics} & 89.61 & 81.65 & \hlb{6.46} & \hlb{0.49}\\
                             & RegNet+$\mathcal{L}_2$  & 89.60  & 81.68 & \hlb{8.00} & \hlb{0.54} \\
                             & RegNet+BIM (ours) & 90.32 & 82.70 & \hlb{5.51} & \hlb{0.37}\\ 
                             & \textbf{RegNet+BIM+$\mathcal{L}_{\text{seg}}$} (ours) & \textbf{92.77} & \textbf{86.76} & \textbf{\hlb{4.39}} & \textbf{\hlb{0.19}} \\ \dmidrule 
    \multirow{7}{*}{RV} & affine only & 81.08 & 70.06 & \hlb{14.31} & \hlb{2.21}\\
                             & B-splines \cite{rueckert1999nonrigid} & 83.59 & 74.33 & \textbf{\hlb{13.70}} & \hlb{2.43} \\
                             & optical flow \cite{perez2013tv} & 84.96 & 75.94 & \textbf{\hlb{13.70}} & \textbf{\hlb{2.15}} \\ 
                             & RegNet+VAE \cite{qin2020biomechanics} & 84.65 & 75.58 & \hlb{16.48} & \hlb{2.31}\\
                             & RegNet+$\mathcal{L}_2$ & 84.80  & 75.73 & \hlb{14.75} & \hlb{2.25}\\
                             & RegNet+BIM (ours) & 85.07 & 76.16  & \hlb{14.55} & \hlb{2.22}\\ 
                             & \textbf{RegNet+BIM+$\mathcal{L}_{\text{seg}}$} (ours) & \textbf{85.93} & \textbf{77.54} & \hlb{14.07} & \textbf{\hlb{2.15}}\\ \dmidrule
    \multirow{7}{*}{Epi}& affine only & 85.07 & 74.50 & \hlb{8.61} & \hlb{0.93}\\ 
                             & B-splines \cite{rueckert1999nonrigid} & 92.33 & 86.12 & \hlb{5.97} & \hlb{0.33} \\
                             & optical flow \cite{perez2013tv} & 92.02 & 85.64 & \hlb{6.09} & \hlb{0.39}\\ 
                             & RegNet+VAE \cite{qin2020biomechanics} & 91.45 & 84.83 & \hlb{7.57} & \hlb{0.48}\\
                             & RegNet+$\mathcal{L}_2$ & 89.92  & 82.22 & \hlb{7.77} & \hlb{0.59}\\
                             & RegNet+BIM (ours) & 91.53 & 84.75 & \hlb{6.20} & \hlb{0.40}\\ 
                             & \textbf{RegNet+BIM+$\mathcal{L}_{\text{seg}}$} (ours) & \textbf{92.57} & \textbf{86.48} & \textbf{\hlb{5.54}} & \textbf{\hlb{0.32}} \\\dbottomrule
    \end{tabular}
    \label{tab:ACDC17}
    \end{adjustbox}
\end{table}

A quantitative assessment of our proposed methods on the ACDC dataset is reported in Table~\ref{tab:ACDC17}. \hlb{Our proposed methods outperform other methods in cardiac structures including the LV and epicardium across all evaluation metrics as well as the Jacobian determinant evaluation shown in Table~\ref{tab:detjac}. Our methods yield higher conformance in terms of DC and JD in RV and similar performance with B-splines and optical flow in terms of HD and ASD.} A quantitative comparison of our proposed method with the others on the LV quantification dataset is included in the supplement  (Table ~\ref{tab:LVQuant19}).

We conduct paired t-tests for statistical significance analysis of our proposed method (RegNet+BIM+$\mathcal{L}_{\text{seg}}$) with the others. Statistically significant improvement ($p<0.05$) is observed for all other methods except for B-splines ($p=0.09$), which might be due to the limited size of the testing set. The complete table is included in the supplementary material (Table~\ref{tab:ACDC17_pvalue}).

\begin{table}[htbp]
    \centering
    \caption{Mean Jacobian determinant comparisons 
    on the ACDC 2017 dataset \cite{bernard2018deep}.}
    \begin{adjustbox}{scale=0.9}
    \begin{tabular}{lc} \dtoprule
          Method & $|\text{det}(J(\hat{u}))-1)|$ \\ \midrule
          B-splines \cite{rueckert1999nonrigid} & 0.2489 \hlb{$\pm$ 0.2735} \\
          optical flow \cite{perez2013tv} & 0.1283 \hlb{$\pm$ 0.3432}\\ 
          RegNet+VAE \cite{qin2020biomechanics} & 0.0088 \hlb{$\pm$ 0.0100} \\
          RegNet+$\mathcal{L}_2$ & 0.0038 \hlb{$\pm$ 0.0051}\\
          \textbf{RegNet+BIM} (ours) & \textbf{0.0035 \hlb{$\pm$ 0.0036}}\\
          RegNet+BIM+$\mathcal{L}_{\text{seg}}$ (ours) & 0.0037 \hlb{$\pm$ 0.0038}\\
  \dbottomrule
  \end{tabular}
  \label{tab:detjac}
  \end{adjustbox}
\end{table}

\subsection{Effect of hyperparameters}
To investigate the impact of our proposed biomechanically-informed regularization and the auxiliary segmentation loss, we vary the weights for each loss term as $\lambda=\{0.005, 0.05, 0.5\}$ and $\gamma=\{0.001, 0.01, 0.1\}$ and report corresponding LV segmentation performance as shown in Table~\ref{tab:hyper_opt}. We also evaluate the effect of Poisson ratio by varying $\nu_p=\{0.35, 0.4, 0.45\}$ in our proposed BIM term. 
From Table~\ref{tab:hyper_opt}, we can see that the $\nu_p$ in the vicinity of $0.4$ has little influence on the segmentation conformance. By increasing $\lambda$, we can see that the performance gradually improves from $0.005$ to $0.05$ and deteriorates quite significantly from $0.05$ to $0.5$. This performance loss is likely due to an underestimate of the DVF to avoid a large penalty when trained with a stronger regularization. By increasing $\gamma$, we observe a steady increase in segmentation performance. However, a stronger auxiliary segmentation regularization might yield unrealistic deformation estimates. The structural similarity index between the fixed and moved images for $\gamma = 0.01$ to $0.1$ also drops from $0.70$ to $0.65$. We also perform an ablation study to illustrate the effectiveness for each term and the detailed statistics on both datasets are included in the supplementary material (Table~\ref{tab:ablation}).

\renewcommand{\arraystretch}{1.2}
\begin{table}[htbp]
    \centering
    \caption{Effect of hyperparameters in RegNet+BIM+$\mathcal{L}_{\text{seg}}$ on ACDC 2017 dataset \cite{bernard2018deep}. $\nu_p$, $\lambda$, and $\gamma$ denote Poisson ratio, weight for BIM regularization
    , and weight for $\mathcal{L}_{\text{seg}}$, respectively. By varying each hyperparameter, we report left ventricle segmentation performance.}
    \begin{adjustbox}{scale=0.9}
    \begin{tabular}{lccc|ccc|ccc}
    \ddtoprule
       Parameter                    & \multicolumn{3}{c|}{$\nu_p$} & \multicolumn{3}{c|}{$\lambda$} & \multicolumn{3}{c}{$\gamma$}  \\\hline
       Value   & 0.35 & 0.4 & 0.45 & 0.005 & 0.05 & 0.5 & 0.001 & 0.01 & 0.1 \\ \ddmidrule
       Dice[\%]$\uparrow$ &  92.74 & 92.77 & 92.66 & 91.51 & 92.77 & 81.03 & 91.14 & 92.77 & 93.27\\
       Jaccard[\%]$\uparrow$ & 86.65 & 86.67 & 86.56 & 84.80 & 86.67 & 68.72 & 84.04 & 86.67 & 87.54\\
       \hlb{HD[mm]}$\downarrow$ & \hlb{4.21} & \hlb{4.39} & \hlb{4.51} & \hlb{7.57} & \hlb{4.39}& \hlb{8.25} & \hlb{5.33} & \hlb{4.39} & \hlb{4.17}\\
       \hlb{ASD[mm]}$\downarrow$ & \hlb{0.19} & \hlb{0.19} & \hlb{0.22} & \hlb{0.35} & \hlb{0.19} & \hlb{1.04} & \hlb{0.33} & \hlb{0.19} & \hlb{0.14}\\
    \dbottomrule
    \end{tabular}
    \end{adjustbox}
    \label{tab:hyper_opt}
\end{table}
\renewcommand{\arraystretch}{1}

\section{Discussion}
\begin{figure}[htbp]
    \centering
    \includegraphics[scale=0.35]{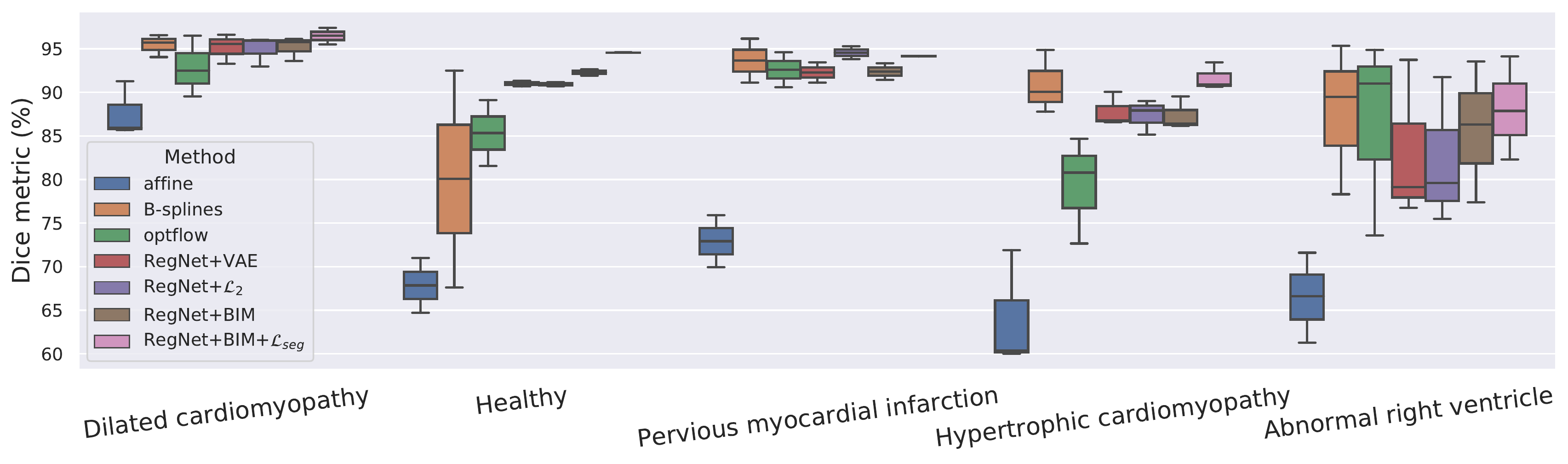}
    \caption{Segmentation conformance in terms of the Dice metric for patients with different pathology from the ACDC dataset \cite{bernard2018deep}.}
    \label{fig:ACDC_disease_bd}
\end{figure}

To analyze the potential registration performance among methods for people with different cardiac conditions, we show the Dice score of each method for patients under different pathologies from the ACDC dataset in the boxplot (Fig.~\ref{fig:ACDC_disease_bd}). From the figure, we can see that our proposed methods yield higher segmentation conformance for dilated cardiomyopathy patients and healthy people. The performance of our proposed methods are similar to others for patients with previous myocardial infarction and hypertrophic cardiomyopathy. Optical flow seems to outperform our proposed method for patients who have abnormal right ventricles. Currently, we validate our proposed methods in the same 2D setting with other competing methods \cite{perez2013tv,qin2020biomechanics}. 

\section{Conclusion}
In this paper, we propose a data-driven approach for cardiac motion estimation with a novel biomechanics-informed modeling regularization loss. Our proposed methods generate more biomechanically plausible DVFs for all cardiac structures in the image without \hlb{requiring myocardium segmentation} or introducing additional steps compared to network-based regularizers. Our proposed methods outperform other state-of-the-art methods on both the ACDC 2017 dataset and the LV quantification 2019 dataset using quantitative evaluation metrics. 
\hlb{Future work will validate current work on other modalities such as tagged MRI and show its effectiveness in more standardized 3D clinical datasets with additional metrics to assess local abnormalities.}

%
%
%
\bibliographystyle{splncs04}
\bibliography{ref}

\section*{Appendix}
\begin{figure}[htbp]
    \centering
    \includegraphics[scale=0.175]{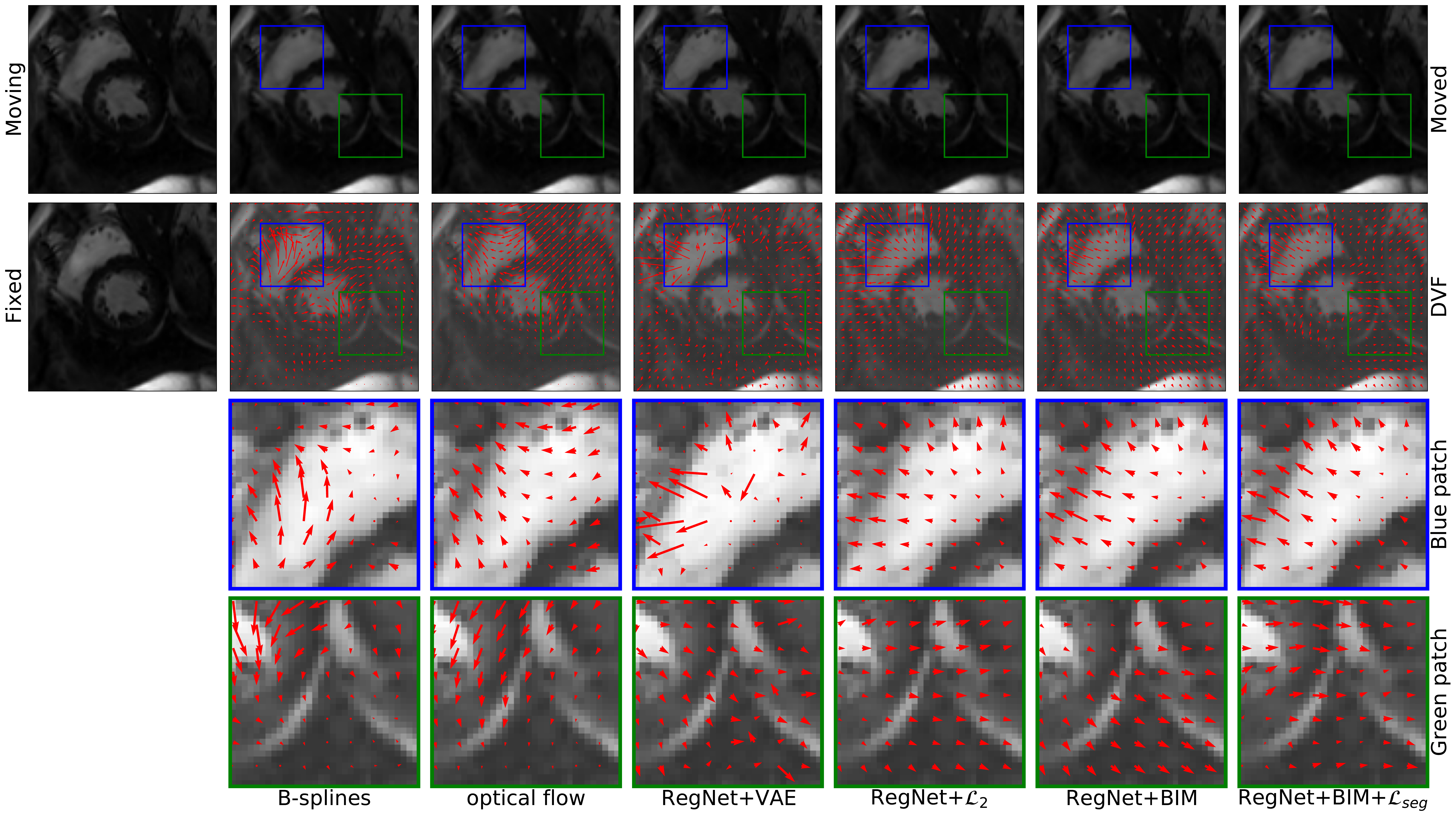}
    \caption{Visual assessment of registration performance 
    on the LV quantification 2019 dataset \cite{xue2018full}. 
    }
    \label{fig:LVQuant_visual}
\end{figure}

\begin{figure}[htbp]
    \centering
    \includegraphics[scale=0.42]{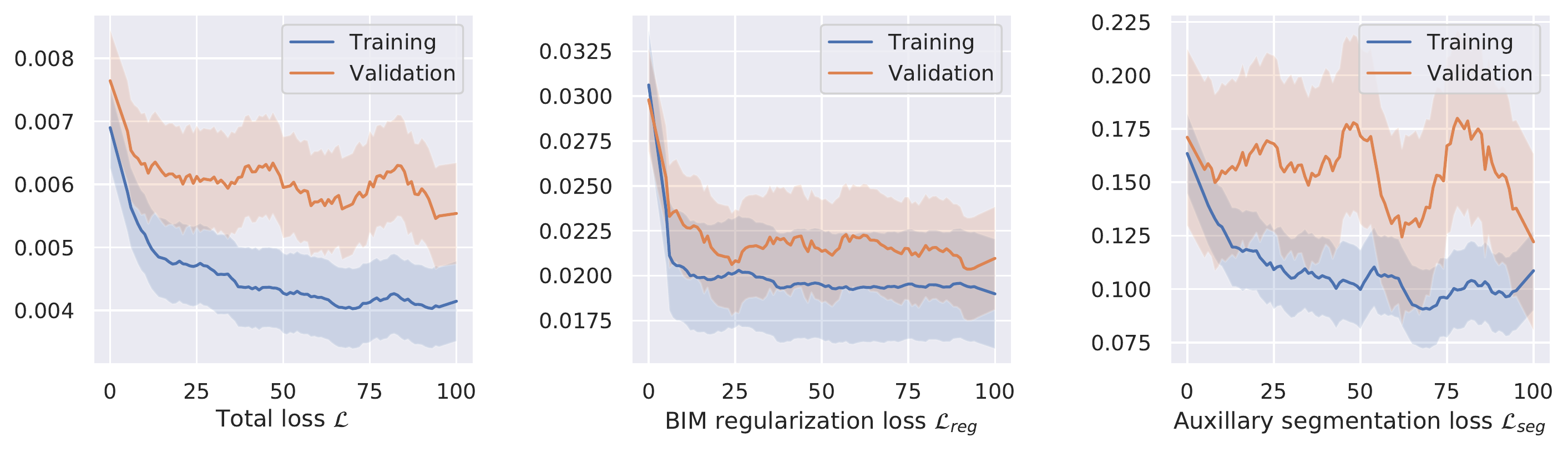}
    \caption{Curves of total loss $\mathcal{L}$, BIM regularization loss $\mathcal{L}_{\text{reg}}$, and auxiliary segmentation loss $\mathcal{L}_{\text{seg}}$ in training and validation, respectively.}
    \label{fig:loss_curve}
\end{figure}

\begin{table}[htbp]
    \centering
    \begin{adjustbox}{scale=0.845}
    \caption{Analysis of paired t-test for statistical significance between RegNet+BIM+$\mathcal{L}_{\text{seg}}$ (ours) and other methods on the ACDC 2017 dataset \cite{bernard2018deep}.}
    \begin{tabular}{lccccc}
    \dtoprule
    Method  & Dice[\%]$\uparrow$    & Jaccard[\%]$\uparrow$    & HD[mm]$\downarrow$      & ASD[mm]$\downarrow$\\\midrule
    affine only & $p<$0.05 & $p<$0.05 & $p<$0.05 & $p<$0.05\\
    B-splines \cite{rueckert1999nonrigid} & $p=$0.09 & $p=$0.09 & $p=$0.09 & $p=$0.06 \\
    optical flow \cite{perez2013tv} & $p<$0.05 & $p<$0.05 & $p<$0.05 & $p<$0.05\\ 
    RegNet+VAE \cite{qin2020biomechanics} & $p<$0.05 & $p<$0.05 & $p<$0.05 & $p<$0.05\\
    RegNet+$\mathcal{L}_2$  & $p<$0.05  & $p<$0.05 & $p<$0.05 & $p<$0.05 \\\dbottomrule
    \end{tabular}
    \label{tab:ACDC17_pvalue}
    \end{adjustbox}
\end{table}

\begin{table}[htbp]
    \centering
    \caption{Quantitative assessment of registration performance on the LV quantification 2019 dataset \cite{bernard2018deep}.}
    \begin{adjustbox}{scale=0.845}
    \begin{tabular}{llcccc}
    \dtoprule
                      &   Method  & Dice[\%]$\uparrow$    & Jaccard[\%]$\uparrow$     & \hlb{HD[mm]}$\downarrow$      & \hlb{ASD[mm]}$\downarrow$\\ \midrule
         \multirow{7}{*}{Endo} & affine only & 82.36 & 71.14 & \hlb{6.92} & \hlb{0.54}  \\ 
                             & B-splines \cite{rueckert1999nonrigid} & 84.54  & 74.30 & \textbf{\hlb{6.31}} & \hlb{0.47}\\
                             & optical flow \cite{perez2013tv} & 83.65 & 73.01 & \hlb{6.52} & \hlb{0.48}\\ 
                             & RegNet+VAE \cite{qin2020biomechanics} & 82.70  & 71.62 & \hlb{6.81} & \hlb{0.52}\\
                             & RegNet+$\mathcal{L}_2$ & 82.49 & 71.33 & \hlb{6.91} & \hlb{0.53} \\
                             & RegNet+BIM (ours) & 84.24 & 73.87 & \hlb{6.63} & \hlb{0.48}\\ 
                             & \textbf{RegNet+BIM+$\mathcal{L}_{\text{seg}}$} (ours) & \textbf{86.65} & \textbf{77.48} & \hlb{6.45} & \textbf{\hlb{0.42}} \\\dmidrule
         \multirow{7}{*}{Epi}& affine only &  93.72 & 88.26 & \hlb{4.62} & \hlb{0.13}  \\ 
                             & B-splines \cite{rueckert1999nonrigid} & 94.19 & 89.10 & \textbf{\hlb{4.27}} & \hlb{0.12} \\
                             & optical flow \cite{perez2013tv} & 94.08 & 88.90 & \hlb{4.33} & \hlb{0.12} \\ 
                             & RegNet+VAE \cite{qin2020biomechanics} & 93.74 & 88.29 & \hlb{4.64} & \hlb{0.13}\\
                             & RegNet+$\mathcal{L}_2$ & 93.72 & 88.26 & \hlb{4.62} & \hlb{0.13} \\
                             & RegNet+BIM (ours) & 93.80 & 88.41 & \hlb{4.71} & \hlb{0.13} \\
                             & \textbf{RegNet+BIM+$\mathcal{L}_{\text{seg}}$} (ours) & \textbf{94.29} & \textbf{89.26} & \hlb{4.50} & \textbf{\hlb{0.11}}\\ \dbottomrule
        \end{tabular}
        \label{tab:LVQuant19}
        \end{adjustbox}
\end{table}


\begin{table}[htbp]
        \centering
        \caption{Ablation study of proposed methods in ACDC 2017 dataset \cite{bernard2018deep} and LV quantification 2019 dataset \cite{xue2018full}.}
        \begin{adjustbox}{scale=0.845}
        \begin{tabular}{llcccc}
        \dtoprule
                      &   Method  & Dice[\%]$\uparrow$    & Jaccard[\%]$\uparrow$     & \hlb{HD[mm]}$\downarrow$      & \hlb{ASD[mm]}$\downarrow$\\\dmidrule
         \multicolumn{6}{c}{ACDC 2017 Dataset \cite{bernard2018deep}} \\\dmidrule
         \multirow{2}{*}{LV} & RegNet  & 82.06  & 70.68 & \hlb{15.45} & \hlb{1.84}\\
                             & RegNet+BIM (ours)  & 90.32 & 82.70 & \hlb{5.51} & \hlb{0.37}\\ 
                             & \textbf{RegNet+BIM+$\mathcal{L}_{\text{seg}}$} (ours) & \textbf{92.77} & \textbf{86.76} & \textbf{\hlb{4.39}} & \textbf{\hlb{0.19}} \\ \midrule
         \multirow{2}{*}{RV} & RegNet  & 84.29  & 75.14 & \hlb{17.15} & \hlb{2.40}\\
                             & RegNet+BIM (ours) & 85.07 & 76.16  & \hlb{14.55} & \hlb{2.22}\\ 
                             & \textbf{RegNet+BIM+$\mathcal{L}_{\text{seg}}$} (ours) & \textbf{85.93} & \textbf{77.54} & \textbf{\hlb{14.07}} & \textbf{\hlb{2.15}} \\ \midrule
         \multirow{2}{*}{Epi}  & RegNet  & 86.67  & 77.06 & \hlb{10.60} & \hlb{0.83}\\
                             & RegNet+BIM (ours) & 91.53 & 84.75 & \hlb{6.20} & \hlb{0.40}\\ 
                             & \textbf{RegNet+BIM+$\mathcal{L}_{\text{seg}}$} (ours) & \textbf{92.57} & \textbf{86.48} & \textbf{\hlb{5.54}} & \textbf{\hlb{0.32}}\\ \dmidrule
         \multicolumn{6}{c}{LV Quantification 2019 Dataset \cite{xue2018full}} \\\dmidrule
         \multirow{2}{*}{Endo} & RegNet  & 84.03 & 73.58 & \hlb{6.76} & \hlb{0.48}\\
                             & RegNet+BIM (ours) & 84.24 & 73.87 & \hlb{6.63} & \hlb{0.48}\\ 
                             & \textbf{RegNet+BIM+$\mathcal{L}_{\text{seg}}$} (ours) & \textbf{86.65} & \textbf{77.48} & \textbf{\hlb{6.45}} & \textbf{\hlb{0.42}} \\ \midrule
         \multirow{2}{*}{Epi} & RegNet  & 93.71  & 88.25 & \hlb{4.70} & \hlb{0.13}\\
                             & RegNet+BIM (ours) & 93.80 & 88.41 & \hlb{4.71} & \hlb{0.13} \\
                             & \textbf{RegNet+BIM+$\mathcal{L}_{\text{seg}}$} (ours) & \textbf{94.29} & \textbf{89.26} & \textbf{\hlb{4.50}} & \textbf{\hlb{0.11}}\\ \bottomrule
        \end{tabular}
        %
        \label{tab:ablation}
        \end{adjustbox}
    \end{table}

\end{document}